\documentclass[12pt]{article} 
\usepackage{epsfig}
\advance\textheight by \topskip
\displaywidth\textwidth %
\linewidth\textwidth %

\begin{document}
\title{Automatic cross-talk removal from multi-channel data}

\author{Bruce Allen, Wensheng Hua, 
\thanks{B.~Allen and W.~Hua are at the Department of Physics,
University of Wisconsin -- Milwaukee,
PO Box 413, Milwaukee, WI 53211, USA. E-mail:
ballen@dirac.phys.uwm.edu.}
\ Adrian C. Ottewill
\thanks{
A.~Ottewill is at the Department of Mathematical
Physics, University College Dublin, 
Belfield, Dublin~4, Ireland.
E-mail: ottewill@relativity.ucd.ie.}}
\maketitle

\begin{abstract}
A technique is described for removing interference from a signal of
interest (``channel 1") which is one of a set of $N$ time-domain
instrumental signals (``channels 1 to $N$").  We assume that channel 1
is a linear combination of ``true" signal plus noise, and that the
``true" signal is not correlated with the noise.  We also assume that
part of this noise is produced, in a poorly-understood way, by the
environment, and that the environment is monitored by channels $2$ to
$N$.  Finally, we assume that the contribution of channel $n$ to
channel $1$ is described by an (unknown!) linear transfer function
$R_n(t-t')$.  Our technique estimates the $R_i$ and provides a way to
subtract the environmental contamination from channel 1, giving an
estimate of the ``true" signal which minimizes its variance.  It also
provides some insights into how the environment is contaminating the
signal of interest.  The method is illustrated with data from a
prototype interferometric gravitational-wave detector, in which the
channel of interest (differential displacement) is heavily contaminated
by environmental noise (magnetic and seismic noise) and laser frequency
noise but where the coupling between these signals is not known in
advance.
\end{abstract}


\section{Introduction}
There are many situations of interest in which data are contaminated by
the environment.  Often this contamination is understood, and by
monitoring the environment it is possible to ``clean up" or ``reduce"
the data, by subtracting the effects of the environment from the signal
or signals of interest.  Examples include measurements of the earth's
magnetic field contaminated by harmonics of $60 \> \rm Hz$, or a
telephone conversation carried on a transmission line, which has been
corrupted by electro-magnetic cross-talk from nearby lines.  The work
in this paper was motivated by another example: the data stream from an
interferometric gravitational radiation detector \cite{ligo}.  In
this instance, the signal of interest is the differential displacement
of suspended test masses.  A small part of this displacement arises
from gravitational waves, but there are also large contributions
arising from contaminating sources, such as the shaking of the optical
tables (seismic noise) and forces due to ambient environmental magnetic
fields.  Particularly at low frequencies, these types of ambient
environmental noise are the fundamental effects limiting the
sensitivity of the instrument \cite{saulsonbook}.  The key point here
is that the gravitational waves are not correlated with any of these
environmental artifacts.

In many such situations, it is possible to monitor the environment,
offering the hope of removing from the signal of interest the
contaminating effects of the environment.  For the prototype
gravitational wave detector used as an example in this paper
\cite{CIT40-meter}, about a dozen of these environmental signals were
monitored, including components of the magnetic field, acoustic
pressure, acceleration of the optical suspension, and so on
\cite{GRASP}. It is not hard to see that in many cases, these
environmental fields add directly into the signal of interest, after
convolution with some (unknown) response function.  For example the
suspension of the optical elements of the interferometer may be
physically modeled by a coupled set of masses, springs, and frictional
elements (dashpots), and thus acts as a mechanical filtering device.
The displacement of the ground is filtered through this suspension and
the resulting displacement is added into the one arising from any
gravitational waves.  Thus if the ground displacement were monitored,
and if we knew the exact transfer function of the suspension, we could
remove from the differential displacement signal the part due to ground
motion.

The difficulty here is that these transfer functions are not known, and
can not be accurately calculated from first principles.  For example the
mechanical filters which isolate the suspension from the ground contain
non-ideal springs, damping elements whose restoring forces are not
proportional to velocity, and so on.  It might in principle be possible
to measure these transfer functions (for example by shaking the ground
in a controlled way) but in many cases this is not practical.

\section{Notation}

Although our methods could be generalized to the case of
continuous-in-time signals, we will assume from here on that all the
signals are discretely sampled in time.  We will assume that the raw
data (channels $1$ to $N$) are time series, sampled at regular
intervals $\Delta t$.  We do {\it not} assume that these sample rates
are the same for all the channels, so in particular $(\Delta t)_n$ will
denote the sample rate of the $n$'th channel.  The $M_n$ (assumed even)
different sample values of channel $n$ at regular time intervals will
be denoted by
\begin{eqnarray}
  Y_n(j) & = & {\rm value\ of\ channel \>} n {\rm \  at\ time\ }t=j (\Delta t)_n \\
\nonumber
& & {\rm \ for\ } j=0,\cdots,M_n-1.
\end{eqnarray}
We assume that each of the channels has been sampled over the entire
time interval $t \in [0,T]$ and thus that $T=M_n (\Delta t)_n$ has the
same invariant value for all channels $n=1,\cdots,N$.
Because the primary goal of our technique is to extract an approximation
of the ``true" or ``uncontaminated" values of channel $1$, we adopt a
special notation for this channel, and use
\begin{equation}
X(j) = Y_1(j)
\end{equation}
to denote the signal of interest. 

Our methods assume that the contamination of channel 1 by the other
channels is described by linear filters or transfer functions.  The
action of a linear filter (convolution in the time domain) is most
simply represented in the frequency domain (where it is just
multiplication), and thus much of our work will take place in the
frequency domain.  The Discrete Fourier Transforms (DFT) of the
channels will be denoted by
\begin{eqnarray}
{\tilde Y}_n(k) & = & \sum_{j=0}^{M_n-1} \exp \left( 2 \pi i  { j k \over M_n}
\right) Y_n(j) \\
\nonumber
& & {\rm \ for\ } k=-M_n/2,\cdots,M_n/2.
\end{eqnarray}
The index $k$ labels frequency bins, and in particular the $k$'th bin
of channel $n$ corresponds to a frequency
\begin{equation}
f_{(n,k)} = k/T.
\end{equation}
Note that throughout this paper, the word {\it band} is used to denote
a collection of adjacent frequency {\it bins}.  We assume that the raw
signals (channels) are real values, i.e. that the $Y_n(k)$ are real,
which implies that ${\tilde Y}_n(k) = {\tilde Y}^*_n(-k)$ where ``$*$"
denotes complex conjugate.

\section{Model (two-channel case)}

We begin by examining the case of only two channels.  This is a good
way to introduce the main ideas of the analysis and the principal
techniques.  In Section~\ref{s:manychannels} we generalize this method
to the $N$-channel case.

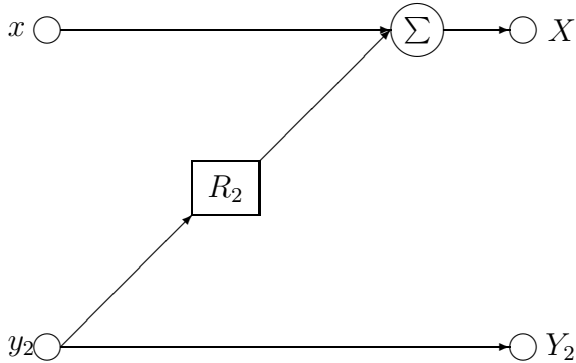
\begin{figure}
\begin{center}
\begin{picture}(240,160)(10,0) 
\put(30,20){\circle{10}}
\put(30,140){\circle{10}}
\put(210,20){\circle{10}}
\put(210,140){\circle{10}}
\put(15,130){\makebox(30,20)[l]{$x$}}
\put(15,10){\makebox(30,20)[l]{$y_2$}}
\put(200,130){\makebox(30,20)[r]{$X$}}
\put(200,10){\makebox(30,20)[r]{$Y_2$}}
\put(85,70){\framebox(25,20){$R_2$}}
\put(155,130){\makebox(30,20){$\sum$}}
\put(170,140){\circle{20}}
\put(35,140){\vector(1,0){125}}
\put(35,20){\vector(1,1){50}}
\put(110,90){\vector(1,1){50}}
\put(35,20){\vector(1,0){170}}
\put(180,140){\vector(1,0){25}}
\end{picture}
\end{center}
\caption{
\label{f:twochannel}
The case where the instrument's output consists of only two channels
($X$ and $Y$) is quite simple.  The ``true" value which the channel of
interest is designed to measure is denoted by $x$.  The actual
instrument output for the channel of interest is denoted by $X$. It is
a linear combination of $x$ and the environmental variable $y_2=Y_2$,
convolved with the response function $R_2$.  By assuming that $x$ and
$y_2$ are not correlated, we can estimate the value of $R_2$ and thus
estimate the value of $x$. }
\end{figure}

Our model may be though of in terms of the diagram in Figure
\ref{f:twochannel}.  The output of the instrument, in other words, the
actual sample values produced by the experiment, are denoted by $X$ and
$Y_2$, in the notation introduced earlier.  These values are to be
thought of as ``imperfect" representations of some true values which
are the variables that the experiment or instrument is attempting to
measure.  However the actual instrumental outputs are not exactly equal
to these values, because they have been contaminated by the
environment.  We denote the true value which the instrument is
attempting to measure by $x$, and the actual output of the instrument
by $X$.  In the two-channel case, the environmental monitor channel is
denoted by $Y_2$; without loss of generality we may assume that it is
equal to the actual value of some environmental variable $y_2$.

In our model, we assume that the true value $x$ of the desired signal
is not present at the output, because it is contaminated by the
environment.  We assume that this contamination may be represented by a
linear filter applied to the environmental variable $y_2$.  For
example, suppose that $x$ is the temperature of a sample of material
(if that material is surrounded by a constant temperature environment)
but in fact the temperature of the environment $Y_2$ is not constant,
and varies with time.  The influence of the environment on the measured
temperature $X$ of the sample is complicated by the fact that the heat
from the environment must diffuse through a thermal insulator before
reaching the sample, so that a change in the temperature of the
environment is not immediately reflected in a change in the temperature
of the sample.  In this example, the effect of the environment on the
sample temperature may be modeled by a first-order linear filter, whose
impulse response decays exponentially in a thermal diffusion time.

In the example that served to motivate this paper, the desired signal
is the difference $x$ in optical phase between two paths of a suspended
interferometer produced by gravitational waves.  However the instrument
contains steering magnets, which are sensitive to the ambient magnetic
fields in the laboratory: these magnetic fields result in forces on the
optical elements which also change the optical phase.  Assuming that
the geometry of the laboratory and of the instrument (which serves to
convert magnetic fields into magnetic gradients) is not changing with
time, one would expect to find a linear filter relationship between
some component of the laboratory magnetic field and the output $X$ of
the relative optical phase channel.  Similar effect arise from seismic
motion and from other sources.

\section{Method (two channel case)}
\label{s:twochannel}
The basic idea of our method is to estimate the transfer functions $R_i$.
This is most easily illustrated in the two-channel case.  The situation
of interest is one in which the transfer function does not change with
time, or changes slowly with time.  This is the case if it is defined
by spring constants (i.e. mechanical coupling) or mutual inductances
(electrical cross-talk) or other quantities that depend upon geometrical
and mechanical properties which change slowly (adiabatically) with time.

To estimate the transfer function requires an averaging process.
It might seem natural to average in time, but the calculations are
easier to understand and express if the averaging is carried out in
frequency space instead.  For this reason, we imagine that the frequency
space occupied by our signal (which for the $n$'th channel is
${\bf R}^{M_n}$) is broken up into subspaces that span frequency bands.
To introduce the notation, we first consider the channel of interest, $X$.
For future convenience we will assume that the Nyquist frequency bin ${\tilde X}(M_1/2)$ does
not contain any useful information (i.e. that an anti-aliasing filter was used
in taking the data) and that we can project our signal onto the
${\bf R}^{M_n-1}$ dimensional subspace that does not include this frequency
bin.
For notational purposes, write this frequency-space representation as the vector
\begin{eqnarray}
{ \tilde {\bf X}} & = & \left[ {\tilde X}(0), {\tilde X}(1),\cdots,{\tilde
                        X}(M_1/2-1) \right] \\
\nonumber
                  & = & \left[ {\tilde {\bf X}^{(0)} }, \cdots, {\tilde
                  {\bf X}^{(B_1-1)} }  \right]
\end{eqnarray}
where we have decomposed the ${\bf R}^{M_1-1}$ into a set of $B_1$
orthogonal vector spaces, each of which contains only the frequencies
in a particular band $b=0,\cdots,B_1-1$.  The number of individual
frequency ``bins" contained in one of the frequency bands is (the
dimensionless integer) $F$, and $ B_n F  = M_n/2$ for $n=1,\cdots,N$.
The number of frequency bands $M_n/2F$ {\it does} depend upon the
channel number (or sample rate) but the number of bins $F$ in a given
band does not.  Thus, the vector ${\tilde {\bf X}^{(0)} } = \left[
{\tilde X}(0), {\tilde X}(1),\cdots,{\tilde X}(F-1) \right]$. In
general, the vector associated with frequency band $b$ consists of
${\tilde {\bf X}^{(b)} } = \left[ {\tilde X}(b F ), {\tilde
X}(bF+1),\cdots,{\tilde X}((b+1)F-1) \right]$. The frequency band
labeled by the dimensionless index $b$ spans a range of physical
frequency $f$ (in cycles/unit-time) given by the half-open interval
\begin{equation}
f \in [f_b ,f_{b+1}) {\rm \ with\ } f_b={b F \over T}.
\end{equation}
Later, we will discuss how we choose the number of frequency bands.
This is related to the question of how much averaging is needed to
accurately estimate the transfer functions.

This notation generalizes in the obvious way to the other channels
$\tilde Y_2,\cdots, \tilde Y_N$.  Note that the number of real degrees of
freedom of the $n$'th channel is $M_n$.  The complex coefficients ${\tilde
Y_n}(i)$ for $i=1,\cdots,M_n/2 - 1$ contain $M_n-2$ of those real degrees
of freedom.  The coefficients ${\tilde Y_n}(0)$ and ${\tilde Y_n}(M_n/2)$
are both real and contain the remaining two real degrees of freedom.
As before (with no significant loss of generality) we will assume that
${\tilde Y_n}(M_n/2)$ is zero, because an anti-aliasing filter has
eliminated any signal contributions near the Nyquist frequency.

To express the correlation between two channels (or the auto-correlation of
a channel with itself) it is useful to introduce a bi-linear inner product.
This is defined by
\begin{equation}
\label{e:normdef}
\left( { \tilde {\bf Y}}^{(b)}_{n_1}  , { \tilde {\bf Y }}^{(b)}_{n_2} \right)
\equiv
\sum_{bF \le k < (b+1)F}  { \tilde { Y}_{n_1}(k)}{ \tilde { Y}_{n_2}^*(k) }.
\end{equation}
This is just the ordinary Cartesian inner product between the two
vectors, after they have been projected into the subspace spanned by
the $b$'th frequency band.  The quantity $\left( { \tilde {\bf
X}}^{(b)}  , { \tilde {\bf X }}^{(b)} \right)$ is the power spectrum of
channel $X$, summed over the $b$'th frequency band: the total power in
the $b$'th frequency band.  Notice that the inner product is {\it only}
defined if both channels $n_1$ and $n_2$ are sampled quickly enough so
that both of them extend up to the $b$'th frequency band.  If the
frequency band lies above the Nyquist frequency of either channel, the
inner product is not defined.  Note also that we could define another
inner product, which is the ordinary Cartesian one (with no projection)
by summing $\left( { \tilde {\bf Y}}^{(b)}_{n_1}  , { \tilde {\bf Y
}}^{(b)}_{n_2} \right)$ over the range $b=0,\cdots,B_{\rm
min}=\min(B_{n_1},B_{n_2})$, but this is used so little that it's not
worth the trouble.

We are now prepared to estimate the transfer function $\tilde R_2(f)$
shown in Figure~\ref{f:twochannel}. Our goal in doing this is to estimate
the ``true" channel of interest $x$.  We denote the estimate of this
quantity with an overbar: $\bar x$.  We also use the overbar to denote
our estimates of derived quantities, for example ${\bar {\bf {\tilde x}}}
$.

We assume that $\tilde R_2(f)$ is {\it complex constant} within
each frequency band $b$, in other words that the transfer function
does not vary rapidly over the frequency bandwidth $\Delta f = F/T$.
For notational convenience, let us denote the constant value of $\tilde
R_2(f)$ in a given frequency band by $r^{(b)}$.  Given the transfer
function $r^{(b)}$ within the frequency band, our estimate of the Fourier
transform of the ``true" channel of interest is
\begin{equation}
\bar {\tilde {\bf x}}^{(b)} =  {\tilde {\bf X}}^{(b)} - r^{(b)} {\tilde {\bf Y_2}}^{(b)}.
\end{equation}
{\bf We assume that the best estimate of the transfer
function in the frequency band $b$ is the one that minimizes the norm $N= \left(
{\bar {\tilde {\bf x}}}^{(b)} , {\bar {\tilde {\bf x}}}^{(b)} \right) $.}
Notice that although the vector ${\bar {\tilde {\bf x}}}^{(b)}$ contains $F$
components, our estimated transfer function $r^{(b)}$ is a single complex number,
containing in practice many fewer degrees of freedom than ${\bar {\tilde {\bf x}}}^{(b)}$.  In this way, the value
of the transfer function averages over the different frequency bins within the
band $b$, and thus corresponds to a time average.

To find $r^{(b)}$ we minimize the norm $N = \left(
{\bar {\tilde {\bf x}}}^{(b)} , {\bar {\tilde {\bf x}}}^{(b)} \right) $.
Under an arbitrary variation $\delta r^{(b)}$ one has
\begin{eqnarray}
\nonumber
\delta N & = &  - \left( 
{ \delta r^{(b)} {\tilde {\bf Y_2}}^{(b)}} ,
{{\tilde {\bf X}}^{(b)} - r^{(b)} {\tilde {\bf Y_2}}^{(b)}}
\right) -\\
\nonumber &&\qquad\qquad
\left(
{{\tilde {\bf X}}^{(b)} - r^{(b)} {\tilde {\bf Y_2}}^{(b)}},
{ \delta r^{(b)} {\tilde {\bf Y_2}}^{(b)}} \right) \\
\nonumber 
& = &   -\delta r^{(b)} \left( 
{  {\tilde {\bf Y_2}}^{(b)}} ,
{{\tilde {\bf X}}^{(b)} - r^{(b)} {\tilde {\bf Y_2}}^{(b)}}
\right) - \textrm{CC} \\
& = &   -2 \Re \left[ \delta r^{(b)} \left( 
{  {\tilde {\bf Y_2}}^{(b)}} ,
{{\tilde {\bf X}}^{(b)} - r^{(b)} {\tilde {\bf Y_2}}^{(b)}} 
\right) \right],
\end{eqnarray}
where ``CC" denotes the complex conjugate of the previous term.
In order that $\delta N$ vanish for all choices of the complex
number $\delta r^{(b)}$ the inner product appearing on the
final line must vanish:
\begin{equation}
\left( 
{  {\tilde {\bf Y_2}}^{(b)}} ,
{{\tilde {\bf X}}^{(b)} - r^{(b)} {\tilde {\bf Y_2}}^{(b)}} 
\right) =0.
\end{equation}
The unique solution to this equation gives our best estimate of the
transfer function $\tilde R_2(f)$ in the frequency band $b$ as:
\begin{equation}
\label{e:estimatedr}
r^{(b)} = {
\left( 
{\tilde {\bf X}}^{(b)} ,
{\tilde {\bf Y_2}}^{(b)} 
\right)
\over
\left( 
{\tilde {\bf Y_2}}^{(b)} ,
{\tilde {\bf Y_2}}^{(b)} 
\right)
}
\end{equation}
We note that instead of minimizing the inner product of our estimate of
the true channel of interest independently within each given frequency
band $b$, we could also have minimized the inner product defined as a sum
over all $B_{\rm min}$ frequency bins; this gives the same result since
vectors obtained by projection onto orthogonal subspaces (corresponding
to different frequency bands) have zero inner product.

\begin{figure}
\begin{center}
\epsfig{file=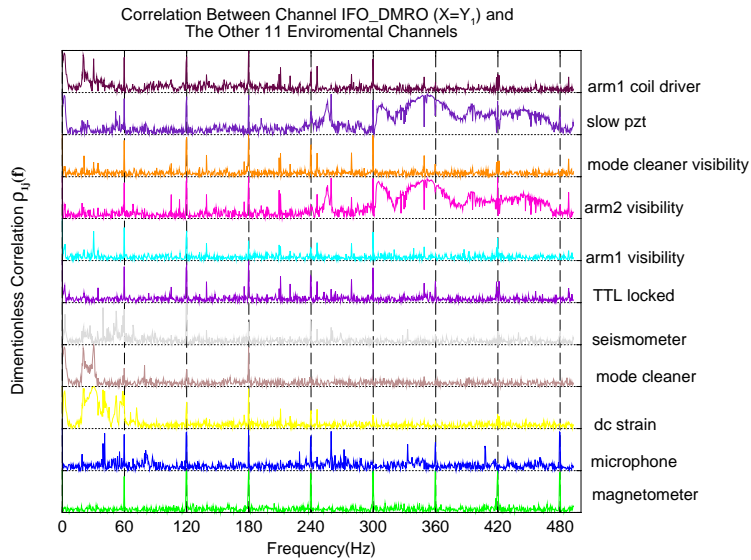,width=7.5truecm,angle=-90}
\caption{ \label{f:correlation}
The estimated correlations between the IFO\_DMRO channel ($X$)
and the other 11 environmental channels. Each individual graph has vertical scale 0 to 1. 
}
\end{center}
\end{figure}

How effective is this procedure likely to be?  Clearly, this depends upon
how much contamination there is, in the channel of interest, and upon how
well the different environmental signals monitor the different sources
of contamination.  In order to quantify these effects, it is useful to
introduce the {\it covariance coefficient} between channels $i$ and $j$
in frequency band $b$, which is defined by
\begin{equation}
\rho_{ij}^{(b)} \equiv \sqrt{
\left| \left( \tilde {\bf Y}_i^{(b)} , \tilde {\bf Y}_j^{(b)} \right) \right|^2
\over
 \left( \tilde {\bf Y}_i^{(b)} , \tilde {\bf Y}_i^{(b)} \right)
 \left( \tilde {\bf Y}_j^{(b)} , \tilde {\bf Y}_j^{(b)} \right)
}.
\label{e:rho}
\end{equation}
From the definition it follows that $ 0 \le \rho_{ij}^{(b)} \le 1$.
This quantity may be interpreted as the (absolute value of) the cosine
of the angle between the vectors representing the $i$'th and $j$'th
channels.  When $\rho_{ij}^{(b)}$ is close to unity this means that the
$i$'th and $j$'th channels are very correlated or anticorrelated; when
close to zero this means that there is no statistically significant
(anti)correlation.  The question ``how large a $\rho_{ij}^{(b)}$ is
statistically significant" will be addressed in
Section~\ref{s:uncorrelated}.  The covariance coefficients
$\rho_{1j}^{(b)}$ between the IFO channel ($X=Y_1$) and the other 11
environmental channels $Y_2,\cdots,Y_{12}$ are shown in
Figure~\ref{f:correlation}.

There are a number of interesting features in the graph of
$\rho_{1j}^{(b)}$ that are worth brief comment.
\begin{itemize}
\item
The magnetometer output shows beautifully strong correlation with the
IFO\_DMRO at all multiples of the line frequency of $60$ Hz.  The large
ambient magnetic fields in the laboratory are probably being produced
by  motors in the ventilation system and transformers in the argon
laser power supply.
\item
The correlations between the microphone and the IFO\_DMRO may reflect
mechanical resonances in the mechanical suspension and isolation systems
which are driven by ambient acoustic noise.
\item
The DC strain is a low-pass filtered version of the IFO\_DMRO channel
of interest: channel $X$!  So in fact it has excellent low-frequency
correlation with the IFO\_DMRO channel because these are measuring
essentially the same thing.  Note that this channel will be left out of
the decontamination procedure that we describe, since that procedure is
intended only for signals that should not show any intrinsic correlation
with the true quantity of interest.
\item
The mode cleaner correlation is easy to understand.  It occurs because
the mode cleaner removes most but not all of the laser frequency
noise.  The remaining frequency noise is converted by the
interferometer into an effective change in the arm length.
\item
The seismometer shows interesting and significant low-frequency
correlation with the IFO\_DMRO.  The mechanical suspension does not
entirely isolate the instrument from ground motions, and these are
subsequently converted into motions of the suspended masses.  These
low-frequency correlations are precisely the sort of correlations that
will be removed by the procedure described in this paper.
\item
The arm 2 visibility and the slow pzt show almost identical correlations
with the IFO\_DMRO channel.  We do not understand why.
\item
The arm 1 coil driver shows very clear low-frequency correlations
with the IFO\_DMRO.  These may be related to the previously-described
correlation between the mode cleaner and the IFO\_DMRO.
\end{itemize}
A technique for simultaneous removal of all of these correlations from
the IFO\_DMRO is described in Section~\ref{s:nchannels}, but for the
moment we return to the simplest, two-channel case.

In the two-channel case, the transfer function in frequency band $b$
was estimated by minimizing the norm $N \equiv  \left( {\bar {\tilde
{\bf x}}}^{(b)} , {\bar {\tilde {\bf x}}}^{(b)} \right)$.  This led to a
unique solution for the estimated transfer function $r^{(b)}$, given by
(\ref{e:estimatedr}).  How much is the norm $N$ reduced when compared
with the corresponding norm of the original channel of interest $ \left(
{ {\tilde {\bf X}}}^{(b)} , { {\tilde {\bf X}}}^{(b)} \right)$ before
any correlations were removed?  This may be found by substituting the
value of $r^{(b)}$ (\ref{e:estimatedr})
into the definition of $N$.  One obtains
\begin{eqnarray}
\nonumber N & \equiv & \left(
{\bar {\tilde {\bf x}}}^{(b)} , {\bar {\tilde {\bf x}}}^{(b)} \right)
\\
\nonumber 
& = &
\left(
{\tilde {\bf X}}^{(b)} - r^{(b)} {\tilde {\bf Y_2}}^{(b)},
{\tilde {\bf X}}^{(b)} - r^{(b)} {\tilde {\bf Y_2}}^{(b)}
\right)
\\ 
\nonumber 
& = &
\left( \tilde {\bf X}^{(b)}, \tilde {\bf X}^{(b)} \right)
- r^{(b)*} \left(\tilde {\bf X}^{(b)}, \tilde {\bf Y_2}^{(b)} \right)
-\\
\nonumber 
&&\qquad
r^{(b)} \left(\tilde {\bf X}^{(b)}, \tilde {\bf Y_2}^{(b)} \right)^*
+|r^{(b)}|^2 \left( \tilde {\bf Y_2}^{(b)}, \tilde {\bf Y_2}^{(b)}\right)
\\
\nonumber 
& = &
\left( \tilde {\bf X}^{(b)}, \tilde {\bf X}^{(b)} \right)
-{
\left|\left( {\tilde {\bf X}}^{(b)} ,{\tilde {\bf Y_2}}^{(b)}\right)\right|^2
\over
\left( {\tilde {\bf Y_2}}^{(b)} ,{\tilde {\bf Y_2}}^{(b)} \right)
}
\\
\nonumber & = &
\left( \tilde {\bf X}^{(b)}, \tilde {\bf X}^{(b)} \right)
\left[
1-{\left| \left( \tilde {\bf X}^{(b)} , \tilde {\bf Y_2}^{(b)} \right) \right|^2
\over
 \left( \tilde {\bf X}^{(b)} , \tilde {\bf X}^{(b)} \right)
 \left( \tilde {\bf Y_2}^{(b)} , \tilde {\bf Y_2}^{(b)} \right)}
\right]
\\
& = &
\left( \tilde {\bf X}^{(b)}, \tilde {\bf X}^{(b)} \right)
\left[1-\left( \rho_{12}^{(b)} \right)^2\right].
\end{eqnarray}
The fractional reduction in the norm $N$ is  $1-\left( \rho_{12}^{(b)}
\right)^2$.  Thus, if an environmental channel is strongly correlated
with the channel of interest, a significant reduction in the norm is
obtained.  As discussed following equation (\ref{e:normdef}) this is may be
though of as a reduction in the total power in the $b$'th frequency band.

\section{An Example (two channel case)}
\label{s:twochanex}

Our example (including Figure~\ref{f:correlation}) is based on data
from the Caltech 40-meter prototype gravitational wave interferometer
\cite{CIT40-meter}.  During one week in November 1994, this instrument
was used to collect data for later analysis.  Between eleven and
fourteen channels of data were collected.  The channel of interest $X$
is the InterFerOmeter Differential Mode Read-out (IFO\_DMRO) and the
other sampled channels consist of environmental and instrumental
monitors.  The channels were sampled at either $9868.42\cdots$ Hz (fast
channels) or at one-tenth that rate (slow channels).

In our first example, we consider only two channels: $X=Y_1$ is the
IFO\_DMRO and $Y_2$ is  IFO\_Mag\_x.  This is the $x-$component of the
magnetic field sampled near one of the optical elements denoted.  Both
of these signals are sampled at the fast rate.  We used $M_1=M_2=10
\times 2048 \times 128$ samples from the 18 November 1994 run 1 data
set, spanning approximately 266 seconds.  To carry out the averaging we
choose $F=128$ frequency bins in each of $B_1=B_2=10\times 2048$
frequency bands.  This is the same data set whose correlations with the
IFO\_DMRO are illustrated in Figure~\ref{f:correlation}.

There is particular reason to believe that the IFO\_DMRO is strongly
contaminated by ambient magnetic field noise (or by signals which are
correlated to that).  This is because the optical elements of the
interferometer suspension are steered and controlled by magnetic
forces.  Many of the optical elements have magnets fastened to them,
and small coils are used to provide some of the servo feedback used to
maintain the optical resonance of the interferometer.  The laboratory
magnetic fields arise from a number of sources, including motors which
are part of the air-circulation systems in the laboratory, and
power-mains and power-supply wiring such as the three-phase current
driving the argon-laser power supply.  It is also possible that ripple
from the power supplies is present in the servo loops whose error
outputs are the source of the IFO\_DMRO signal.

\begin{figure}
\begin{center}
\epsfig{file=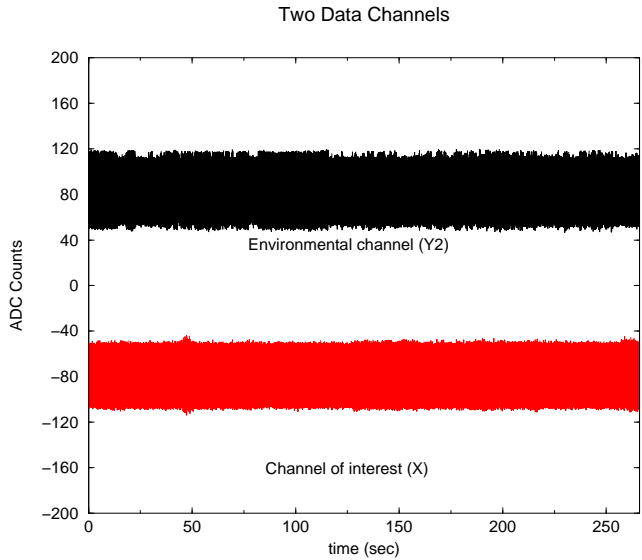,width=7.5truecm,angle=-90}
\caption{ \label{f:timedomain} 
Approximately 266 seconds of two channels of the Caltech 40-meter
interferometer output, after low-pass filtering to remove all
frequency components above $ 0.1 \;f_{\rm Nyquist}$.  $X$ denotes the
Differential Mode Readout, which is the channel of interest.  $Y_2$ is the
output of a magnetometer, sensing a component of the local magnetic field.
These two signals are both contaminated by many harmonics of 60 Hz.
They are shifted by $\pm 80$ ADC counts for clarity.
}
\end{center}
\end{figure}

\begin{figure}
\begin{center}
\epsfig{file=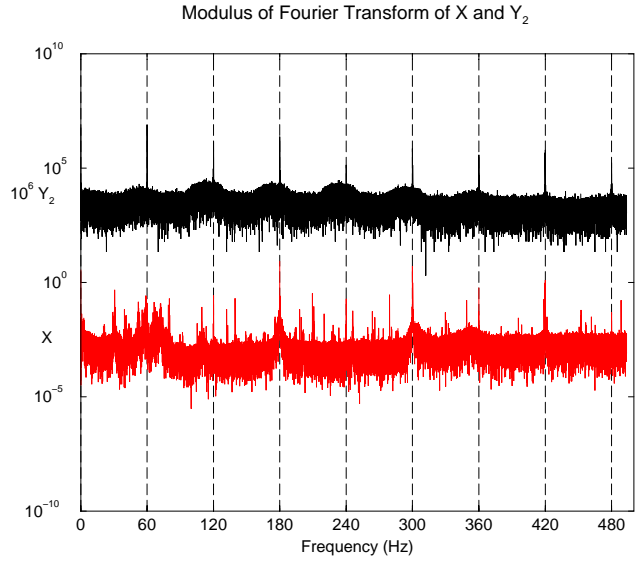,width=7.5truecm,angle=-90}
\caption{ \label{f:freqdomain} 
The amplitude spectrum of the data sets from
Figure~\ref{f:timedomain}. Notice that there are strong line-like
features at the harmonics of 60 Hz, particularly around 180 and 300 Hz
in the channel of interest.  The former may be due to the laser's power
supply producing cross-talk in other electronics. This graph shows only frequencies $\le 0.1 \;f_{\rm Nyquist}$.}
\end{center}
\end{figure}

\begin{figure}
\begin{center}
\epsfig{file=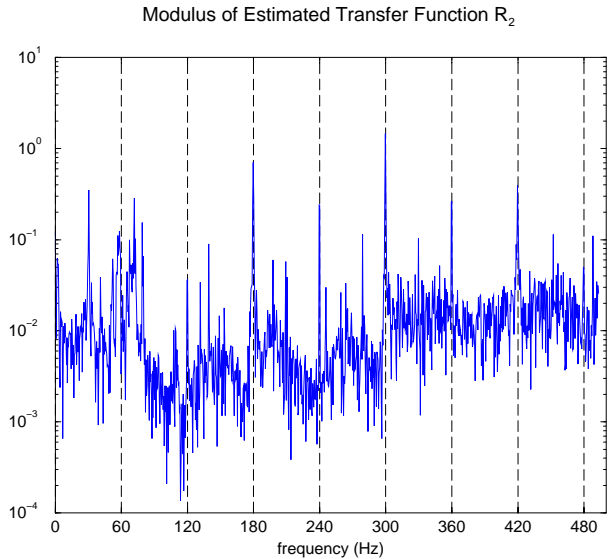,width=7.5truecm,angle=-90}
\caption{ \label{f:coupling}
The estimated coupling function $R_2(f)$ between the IFO channel ($X$)
and the magnetometer ($Y_2$).   This estimate is dominated by noise, except
at frequencies where its modulus is large compared to nearby frequencies.
At these frequencies, the estimate is accurate.   The frequencies at
which $R_2$ can be accurately estimated includes (but is not limited to)
many of the 60 Hz line harmonics.
}
\end{center}
\end{figure}

Figure \ref{f:timedomain} shows the two channels $X$ and $Y_2$ for 266
seconds.  Because our primary goal is to remove low-frequency ($f <
987/2$~Hz) contamination from $X$, these channels have been low-pass
filtered by (1) transforming into the frequency domain (2) setting to
zero all spectral amplitudes at frequencies $\ge 0.1 \;f_{\rm Nyquist}$
then (3) transforming back into the time-domain.  Although it is not
obvious from the graphs, both channels contain strong sinusoidal
components at multiples of the line frequency 60 Hz.
Notice that the rms value of channel $X$ is about 30 ADC counts.
Also notice that the small instrumental feature (blip) around
$t=46$~sec is almost obscured by the surrounding ``hash". The Fourier
transforms of these two channels are shown in
Figure~\ref{f:freqdomain}.
Using the procedure that we have described, we can estimate the coupling
$R_2(f)$ between these two channels.  This is shown in Figure~\ref{f:coupling}.
In each frequency band, the estimate of $R_2$ is a sum over the $F=128$
different frequency bins contained in that band.  If there is no
correlation between the two channels, the expected value of this
sum behaves like a random walk, accumulating proportional to $\sqrt{F}$. (The
case where there is no correlation is considered in detail in
Section~\ref{s:uncorrelated}.)  In frequency bands where the two channels
are correlated, the expected value of the sum accumulates proportional to $F$.

\begin{figure}
\begin{center}
\epsfig{file=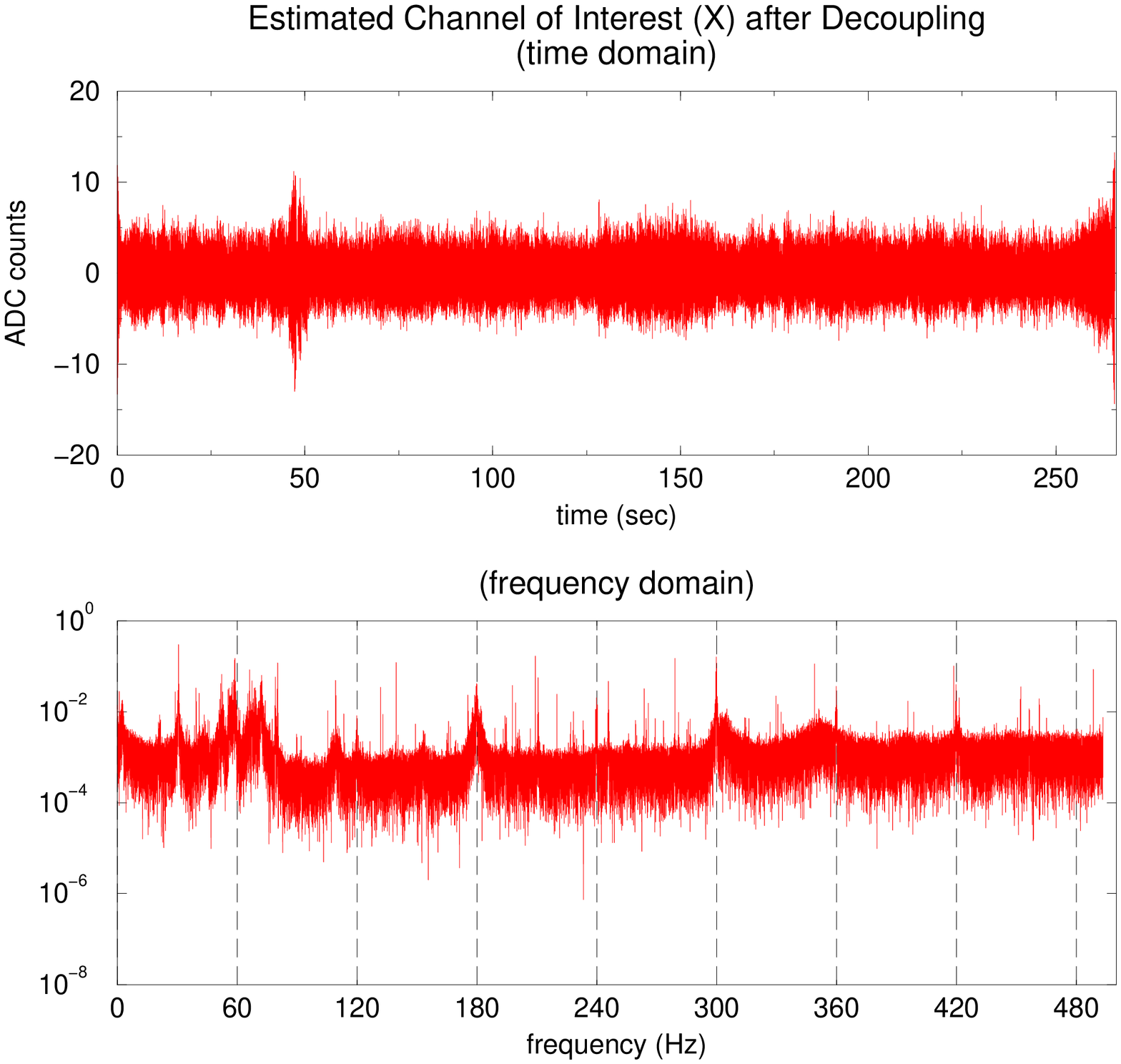,width=7.5truecm,angle=-90}
\caption{ \label{f:cleaned}
The final result of the estimation process that is described is an
estimate of the true value of the IFO\_DMRO channel after subtraction
of the correlated contamination.  This should be compared with the
original time/frequency domain data shown in Figures \ref{f:timedomain}
and \ref{f:freqdomain}.  Notice that the rms of the time-domain signal
has been reduced by about a factor of six!  There has been a significant
reduction in the spectral content of the signal around 180 and 300 Hz,
and the instrumental effect around 46 seconds is much more apparent.
}
\end{center}
\end{figure}

The final result, Figure~\ref{f:cleaned} shows the estimated ``true"
value of the IFO Differential Mode Output channel, after subtracting the
estimated crosstalk.

\section{Method ($N$-channel case)}
\label{s:nchannels}
In Section~\ref{s:twochannel} we showed how it is possible to obtain
an estimate of the coupling between two channels, by searching for the
linear combination (in frequency space) that minimizes the variance of
the channel of interest.  In this section, we generalize this method
to the $N$-channel case, where in addition to the channel of interest,
there are $N-1$ additional environmental monitoring channels.

The basic idea is identical.  We estimate the ``true" value of the
channel of interest as:
\begin{equation}
\label{e:remove}
\bar {\tilde {\bf x}}^{(b)} =  {\tilde {\bf X}}^{(b)} - \sum_{j=2}^N
r^{(b)}_j {\tilde {\bf Y_j}}^{(b)}.
\end{equation}
Here the $r^{(b)}_j$ are a set of $N-1$ coupling constants: they are
our estimates of the contribution that the channel $Y_j$ makes to the
channel of interest $X$ in the frequency band $b$.  As before, we
choose these coupling constants in the way that minimizes the total
power in the channel of interest, assuming that they are constant
throughout the $b$'th frequency band.  This means that we choose the
$r^{(b)}_j$ in order to minimize the expected value of the norm
$N = \left(
{\bar {\tilde {\bf x}}}^{(b)} , {\bar {\tilde {\bf x}}}^{(b)} \right) $,
under an arbitrary variation $\delta r^{(b)}_j$.  
\begin{eqnarray}
\nonumber
\delta N & = &  \sum_{j=2}^N \left( - \left( 
{ \delta r_j^{(b)} {\tilde {\bf Y_j}}^{(b)}} ,
{{\tilde {\bf X}}^{(b)} - \sum_{k=2}^N r_k^{(b)} {\tilde {\bf Y_k}}^{(b)}}
\right) - \right.\\
\nonumber &&\qquad \qquad
\left. \left(
{{\tilde {\bf X}}^{(b)} - \sum_{k=2}^N r_k^{(b)} {\tilde {\bf Y_k}}^{(b)}},
{ \delta r_j^{(b)} {\tilde {\bf Y_j}}^{(b)}} \right) \right) \\
\nonumber 
& = &   
\sum_{j=2}^N \left(
-\delta r_j^{(b)} \left( 
{  {\tilde {\bf Y_j}}^{(b)}} ,
{{\tilde {\bf X}}^{(b)} - \sum_{k=2}^N r_k^{(b)} {\tilde {\bf Y_k}}^{(b)}}
\right) - \textrm{CC}  \right) \\
& = &   -2 \Re \left[ \sum_{j=2}^N \delta r_j^{(b)} \left( 
{  {\tilde {\bf Y_j}}^{(b)}} ,
{{\tilde {\bf X}}^{(b)} - \sum_{k=2}^N r_k^{(b)} {\tilde {\bf Y_k}}^{(b)}} 
\right) \right].
\end{eqnarray}
In order for this quantity to vanish under all variations of the $N-1$ coupling
constants $\delta r_j^{(b)}$  one must satisfy $N-1$ equations (for $j=2,\cdots,N)$ :
\begin{equation}
\label{e:reqns}
\left( 
{  {\tilde {\bf Y_j}}^{(b)}} ,
{{\tilde {\bf X}}^{(b)} - \sum_{k=2}^N r_k^{(b)} {\tilde {\bf Y_k}}^{(b)}} 
\right) =0.
\end{equation}
This may be conveniently written in matrix/vector form.  To do so, define the
{\it correlation matrix estimate in the $b$'th channel}:
\begin{equation}
\label{e:matrixdef}
C^{(b)}_{jk} = \left( {\tilde {\bf Y_j}}^{(b)} , {\tilde {\bf Y_k}}^{(b)} \right).
\end{equation}
This matrix $C^{(b)}_{jk}$ is Hermitian
and positive semi-definite.  Notice that the entries of this correlation matrix are defined for
$j,k=1,\cdots, N$ since by definition the channel of interest $X = Y_1$.
This means that ``intrinsically" $C$ is a square $N\times N$ matrix.

The equations satisfied by the coefficients (\ref{e:reqns}) may now be
written as
\begin{equation}
C^{(b)}_{j1} = \sum_{k=2}^N C^{(b)}_{jk} r_k^{(b)}.
\end{equation}
Notice that the left hand side is determined by the correlations between
the channel of interest and the environmental channels.  The matrix that
appears on the right hand side is determined by the correlation between
the different environmental channels.  In the case where these are not
correlated (i.e., a given environmental channel is only correlated with
itself) then the matrix on the right hand side is diagonal, and the
situation is very similar to the two-channel case.

If all of the channels are non-zero in at least one bin in frequency
band $b$ then the matrix is Hermitian and positive definite, so that
it may be inverted.  We denote the inverse of this matrix by the
symbol $C^{-1}$.  Note: this is {\it not} the inverse of an $N\times N$
matrix.  It is the inverse of the $(N-1) \times (N-1)$ matrix defined by
(\ref{e:matrixdef}) for $j,k=2,\cdots,N$.

The coupling constants that
minimize the variance in the channel of interest are now given by:
\begin{equation}
\label{e:ri}
r_j^{(b)} = \sum_{k=2}^N \left(C^{-1}\right)_{jk}C_{k1} \quad {\rm for\
}j={2,\cdots,N}.
\end{equation}
Although it is tempting to interpret this equation as ``inverse of
a matrix times the matrix" and replace the rhs by $\delta{j1}$ this
is not correct, because $C^{-1}$ is the inverse of an $(N-1) \times
(N-1)$ matrix.

It is again possible to ask how much the norm $N$ is reduced when compared with the corresponding norm of the original channel of interest $ \left(
{ {\tilde {\bf X}}}^{(b)} , { {\tilde {\bf X}}}^{(b)} \right)$ before
any correlations were removed.  This may be found by substituting the
value of $r^{(b)}$ (\ref{e:estimatedr})
into the definition of $N$.  One obtains
\begin{eqnarray}
N & \equiv & \left(
{\bar {\tilde {\bf x}}}^{(b)} , {\bar {\tilde {\bf x}}}^{(b)} \right)
\cr
& = &
\left(
{\tilde {\bf X}}^{(b)}- \sum_{j=2}^N
r^{(b)}_j {\tilde {\bf Y_j}}^{(b)} ,
{\tilde {\bf X}}^{(b)} - \sum_{j=2}^N
r^{(b)}_j {\tilde {\bf Y_j}}^{(b)}
\right)
\cr 
& = &
\left(
{\tilde {\bf X}}^{(b)}- \sum_{j=2}^N
r^{(b)}_j {\tilde {\bf Y_j}}^{(b)},
{\tilde {\bf X}}^{(b)} 
\right)
\cr 
& = &
\left( \tilde {\bf X}^{(b)}, \tilde {\bf X}^{(b)} \right)
\left[1- |\rho^{(b)}|^2\right].
\end{eqnarray}
where the second line follows from Eq.~\ref{e:reqns} and we have
defined 
\begin{eqnarray}
\label{e:rho_n}
|\rho^{(b)}|^2 &=& \sum_{j=2}^N r^{(b)}_j {\left(
{\tilde {\bf Y_j}}^{(b)},{\tilde {\bf X}}^{(b)}\right) \over 
\left ({\tilde {\bf X}}^{(b)},{\tilde {\bf X}}^{(b)}\right) }\cr
&=& \sum_{j=2}^N \sum_{k=2}^N C_{1j}^*\left(C^{-1}\right)_{jk} C_{k1}
/C_{11}^{(b)}.
\end{eqnarray}
The second form expresses $|\rho^{(b)}|^2$ in manifestly positive
definite form, while from its definition it is always less than or
equal to 1.  The quantity $|\rho^{(b)}|^2$ provides a useful measure of
the total improvement in the signal.  To understand which environmental
channels led to this improvement one may study the $N-1$ pairwise
covariance coefficients $|\rho^{(b)}_{1j}|^2$.

\section{Example ($N$-channel case)}
\label{s:manychannels}
Our $N$-channel example is based on the same 18 November 1994 run 1 data
from the Caltech 40-meter prototype gravitational wave interferometer
\cite{CIT40-meter} that was used in the previous 2-channel example in
Section~\ref{s:twochanex}.  As before, the channel of interest $X$ is
the InterFerOmeter Differential Mode Read-out (IFO\_DMRO) and the other
11 sampled channels consist of environmental and instrumental monitors.
Three of the channels (including IFO\_DMRO) were sampled at the fast rate
of $9868.42\cdots$ Hz and the other nine were sampled at exactly one-tenth
that rate.  The different channels are shown in Table~\ref{t:channels}.
The covariance coefficients $\rho_{1j}$ between these channels and the
IFO\_DMRO channel were previously shown in Figure~\ref{f:correlation}.

\begin{table}
\caption{Channel assignments for the November 1994 data runs.  Channels
0-3 are the ``fast" channels, sampled at about 10 kHz; the remaining
twelve are the ``slow" channels, sampled at about 1KHz. Note that the
power stabilizer channel was accidentally disconnected until
approximately 20:00 local time and so was not used by us, and that some
channel numbers were not present in the data.}
\begin{center}
\begin{tabular}[]{|l|l|l|}
\hline
Channel \# & Content & FRAME name\\
\hline
0   & IFO output        & IFO\_DMRO\\
1   & magnetometer      & IFO\_Mag\_x \\
2   & microphone        & IFO\_Mike\\
\hline
4   & dc strain         & IFO\_DCDM \\
5   & mode cleaner pzt  & PSL\_MC\_V \\
6   & seismometer       & IFO\_Seis\_1 \\
7   & slow pzt          & PSL\_SPZT\_V \\
8   & power stabilizer  & PSL\_PSS  \\
10  &  TTL locked       & IFO\_Lock \\
11  & arm 1 visibility  & IFO\_EAT\\
12  &  arm 2 visibility &  IFO\_SAT\\
13  & mode cleaner visibility & IFO\_MCR\\
15  & arm 1 coil driver & SUS\_EE\_Coil\_V \\
\hline
\end{tabular}
\end{center}
\label{t:channels}
\end{table}

\begin{figure}
\begin{center}
\epsfig{file=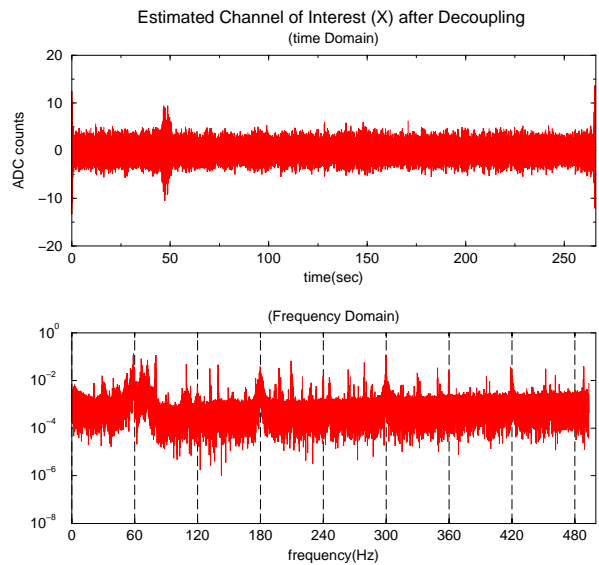,width=7.5truecm,angle=-90}
\caption{ \label{f:cleaned2}
The final result of the estimation process, to remove contamination
from 11 monitored environmental channels from the IFO\_DMRO channel.
This should be compared with the original time/frequency domain data
shown in Figures \ref{f:timedomain} and \ref{f:freqdomain}.
}
\end{center}
\end{figure}

As before, we used the first $M_1=M_2=10 \times 2048 \times 128$
samples from the  data set, covering about 266 seconds.  As before,
to carry out the averaging we choose $F=128$ frequency bins in each of
$B_1=B_2=10\times 2048$ frequency bands.  Because the DC strain channel
is effectively just a low-pass filtered version of the IFO\_DMRO channel,
it was left out of the removal process.  The result of this procedure
is shown in Figure~\ref{f:cleaned2}.
It is very useful to compare
this with the previous 2-channel case, where we removed only contamination
that was correlated with the magnetometer channel.  In comparison
to this previous case, the following features are evident:
\begin{itemize}
\item
The ``end effects" that are apparent in the two channel case (Figure~\ref{f:cleaned})
are no longer present: these were not end effects but contamination of the
IFO\_DMRO channel by an interfering signal.
\item
Comparison of the power spectra
(Figures~\ref{f:freqdomain},\ref{f:cleaned},\ref{f:cleaned2}) in the region
below 60 Hz shows that a significant reduction in low-frequency content
has been obtained.
\item
The time-domain properties of the estimated detector noise are now {\it
more} uniform, rather than less uniform.  This is good evidence that
the signal content which is being removed is in fact a true correlated
signal and not merely an artifact of the subtraction procedure
described here.
\end{itemize}

\section{Reducing the effects of correlated sidelobes}
When we began this work, our original intent had been to carry out a
procedure similar to the one just described.  However that procedure
failed, for reasons that are interesting, and are worth explaining here.

The procedure which failed can be summarized as follows:
\begin{itemize}
\item
Take long stretches of data from each of the $N$ channels, spanning a
time interval of length $T$.
\item
Cut them into $T/\tau$ short segments of length $\tau$ (say, one second long) .
\item
Transform these into the frequency domain.
\item
For each short segment, and in each frequency bin, calculate an $N
\times N$ matrix containing the products of the Fourier amplitudes of
the different channels.
\item
In each frequency bin, average the $T/\tau$ matrices thus obtained to get an
estimate of the correlation matrix.
\item
Use this correlation matrix to estimate the transfer function $R_j$
that minimizes the total power in each frequency band.
\end{itemize}
The reason why this procedure failed is not hard to understand.

One might expect that in this procedure, since the length of each segment
in the time-domain is $\tau$, the frequency-resolution of this method
is $\Delta f \sim \tau^{-1}$.  Thus, for example, the line-frequency
harmonics appearing at multiples of $60 \> \rm Hz$ might be expected to
be resolved within a band about $\pm 1 \> \rm Hz$ about their true locations.
This is correct.

\begin{figure}
\begin{center}
\epsfig{file=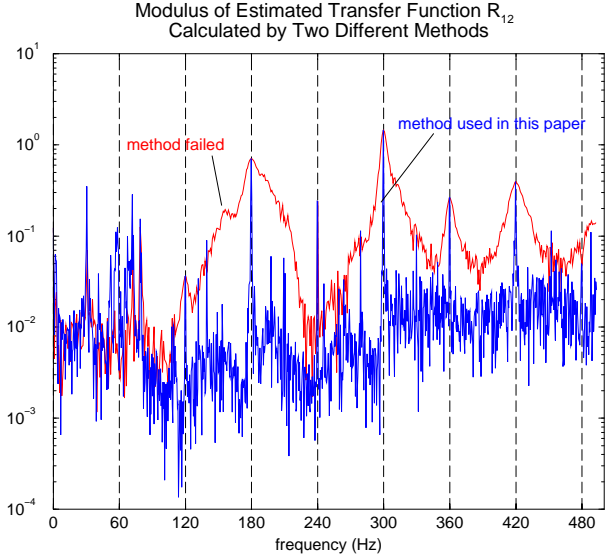,width=7.5truecm,angle=-90}
\caption{ \label{f:correlatedsidelobes}
A comparison of two methods of estimating the transfer function $R_{12}$.
The dark curve shows the method used in this paper: in each frequency
band $b$ the estimate is constructed as an average over nearby frequency
bins.  The light curve shows the method that failed: it is essentially the
time average of a single frequency bin in a sequence of short Fourier
transforms.  It fails because sidelobes of strong line features cause
spurious correlations over a much wider range of frequencies than $\Delta
f = 1/T$.
}
\end{center}
\end{figure}

The problem occurs because in many instances, these line-like features
in the frequency domain have {\it much} larger amplitude (by orders
of magnitude) than the neighboring frequency bins.  In addition, these
line-like features do not lie precisely at the center of a frequency bin
(in the time domain, the corresponding sinusoids do not undergo an integer
number of oscillations during the time-interval $\tau$).  Consequently,
these line-like features exhibit sidelobes of the windowing function.  In
the method that we have described, this windowing function is rectangular
(on or off) but even if a more sophisticated and smoothly-varying window
function is chosen, the sidelobes are still present.  These sidelobes
are much smaller than the central maximum, and depending upon the
choice of window function, they fall off as some (inverse) power of the
separation in frequency bins from the central line.  Since the energy
in the central line is so large compared with neighboring frequencies,
these sidelobes, while insignificant compared with the central line
feature, are still large enough to completely dominate the signals at
neighboring frequencies.  Consequently, one finds that there are large
correlations arising from the central line-like features, extending
out over a range of frequencies that is quite large compared to $\Delta
f \sim \tau^{-1}$.  In many of the instances which we examined, these
correlated sidelobes dominate the true correlation out to $50 \Delta f$.
This is shown in Figure~\ref{f:correlatedsidelobes}.

The failure of this other method may be easily summarized as follows.
Although the energy arising from a sinusoidal signal present in several
channels is largely confined to a (small) bandwidth $\Delta f$, the
correlation arising from this signal can dominate the correlation
over a bandwidth which is fifty times larger!  The resulting loss in
frequency resolution is unacceptable.  For this reason, we don't use
(or recommend!) this method for estimating the correlations between
different channels.

\section{Avoiding False Dismissal of ``Correlations"(two-channel case)}
\label{s:uncorrelated}

The methods that we have described for removing environmental
contamination or crosstalk from signals of interest assumes that there
is no correlation between the environmental monitors and the signal of
interest, and thus that any correlation which {\it is} found is due to
``leakage" or ``crosstalk" in the instrument.  If this assumption is
satisfied, one might well ask, ``Is there a danger of falsely removing
correlations which do not in fact exist in the observed signals?"
To quantify this requires that we make assumptions about the statistics
of any uncorrelated signals.

Suppose that we consider the case where the $N$ channels are
independent uncorrelated Gaussian random variables, with a white power
spectrum.  For simplicity let us also assume that each has zero mean
value and unit variance.  This is a situation where a good technique
for removal of correlated noise from the channel of interest should do
absolutely nothing, since there is no correlation to remove!  How does
the technique described here perform in this situation?

For simplicity, consider first the two-channel case of
Section~\ref{s:twochannel}.  Suppose that the signal values $X(i)$ and
$Y_2(j)$ are independent Gaussian random variables with mean value zero
and unit variance.  In this case, the Fourier amplitudes $\tilde X(i)$
and $\tilde Y_2(j)$ are also independent Gaussian random variables.
The estimated transfer function (\ref{e:estimatedr}) for a particular
frequency band has mean value zero.  The expectation value of its square
is given by
\begin{eqnarray}
\left\langle \left|r^{(b)}\right|^2 \right\rangle & = &
\left\langle
{
 \sum_{i \in (b)} \sum_{j \in (b)}
 \tilde X(i) \tilde Y^*(i)
 \tilde X^*(j) \tilde Y(j)  
\over 
 \sum_{i \in (b)} \sum_{j \in (b)} 
 |\tilde Y(i)|^2 |\tilde Y(j)|^2
}
\right\rangle \cr
& = &
\langle |\tilde X|^2 \rangle 
\left\langle
{
 \sum_{j \in (b)}  |\tilde Y(j) |^2 
\over 
 \sum_{i \in (b)} \sum_{j \in (b)} 
 |\tilde Y(i)|^2 |\tilde Y(j)|^2
}
\right\rangle \cr
& = &
\langle |\tilde X|^2 \rangle 
\left\langle
{
 1 
\over 
 \sum_{i \in (b)}  
 |\tilde Y(i)|^2 }
\right\rangle
\end{eqnarray}
The calculation of the last quantity is slightly complicated and may be
found in Appendix~\ref{s:rho}.  Here, we approximate it in the case
where the frequency band $(b)$ contains many frequency bins.  Since the
number of frequency bins in the $b$'th frequency band is denoted by
$F$, we will assume that $F >> 1$.  In this case one obtains
\begin{equation}
\left\langle \left|r^{(b)}\right|^2 \right\rangle = {1 \over F}.
\end{equation}
Because the estimated transfer function $r$ and the covariance $\rho^2$
are related by
\begin{equation}
 \left( \rho^{(b)}_{1j} \right)^2 = |r_j^{(b)}|^2 { ( \tilde {\bf Y}_j ,   \tilde {\bf Y}_j )
\over ( \tilde {\bf X} ,   \tilde {\bf X} ) },
\end{equation}
our simple example of two uncorrelated channels would have the expectation
value of $\rho_{12}^2 $ in each frequency band equal to $1/F$.  Thus, on the
average, blind application of our method would reduce the variance of the
channel of interest by the fraction
\begin{eqnarray}
\label{e:false-remove}
\nonumber
N & \equiv & \left(
{\bar {\tilde {\bf x}}}^{(b)} , {\bar {\tilde {\bf x}}}^{(b)} \right)
\\
\nonumber
& = &
\left( \tilde {\bf X}^{(b)}, \tilde {\bf X}^{(b)} \right)
\left[1-\left( \rho_{12}^{(b)} \right)^2\right]\\
& = &  
\left( \tilde {\bf X}^{(b)}, \tilde {\bf X}^{(b)} \right)
\left(1 - {1 \over F} \right).
\end{eqnarray}
This is clearly unacceptable since there is no correlation actually
present, and the power in the channel of interest should not be reduced
at all.  In the case where we have two uncorrelated Gaussian random
channels, with for example $F=128$, the direct application of the
method described here will reduce the power in the channel of interest
by almost one percent!

The problem that we are describing is that of incorrectly or falsely
removing correlations that are not really present!  If the length of
the data set were extremely long, so that the number $F$ of frequency
bins in any given frequency band were very large, then this problem
would disappear.  However in practical work, it is unreasonable to have
very large numbers of frequency bins $F$ in each band.

One simple solution to this problem is to threshold on the covariance.
In other words, we examine each environmental channel in turn, and ask
if it is correlated with the channel of interest.  If such a covariance
is present {\it at a statistically-significant level}, the correlation
is removed.  Otherwise, the correlation is not removed.  Since the
expectation value of $\rho^2$ is $1/F$, we can set a threshold of say
$10/F$

\section{Avoiding False Dismissal of ``Correlations"(n-channel case)}
\label{s:thresh_N}

In the N-channel case, there is also a risk of falsely removing
``correlations'' that are not present. In Section\ref{s:nchannels}, we
introduced the correlation matrix by equation (\ref{e:matrixdef}). All
the following calculations are based on that matrix. Each entry of that
matrix, $ C^{(b)}_{jk} = \left( {\tilde {\bf Y_j}}^{(b)} , {\tilde {\bf
Y_k}}^{(b)} \right)$, is the correlation between channels $i$ and $j$
in frequency band $b$. According to Appendix \ref{s:rho}, because the
number $F$ of frequency bins in frequency band $b$ is finite, the
correlation between any two channels can not be calculated precisely.
Consequently there is a risk of finding correlations when none exist,
and then incorrectly removing them.

One method to avoid false dismissal of ``correlation'' is to threshold
on every entry of the correlation matrix, $C^{(b)}_{jk}$. We calculate
the absolute value of the covariance coefficient between channels $j$
and $k$ in frequency band $b$, $\rho_{jk}^{(b)}$, which is defined by
equation (\ref{e:rho}). If $\rho_{jk}^{(b)}$ is smaller than some
threshold value $\rho^*$ (for example $ \rho^*= {10 \over F}$), then we
set the corresponding entry  $C^{(b)}_{jk}$ in the correlation matrix
to zero. If $\rho_{ij}^{(b)}$ is greater than the threshold value
$\rho^*$, then we leave the corresponding entry $C^{(b)}_{jk}$ in
correlation matrix unchanged. We use, $D^{(b)}_{jk}$, to denote the
correlation matrix after thresholding:
\begin{equation}
\label{e:threshmatrix}
D^{(b)}_{jk}= \cases{ C^{(b)}_{jk} & $\textrm{if}\ \rho_{jk}^{(b)}>\rho^*  {\textrm{, or}} $  \cr
&\cr
0 & ${\textrm{otherwise}} \ .$}
\end{equation} 
The next step is to calculate the coupling constants using equation
(\ref{e:ri}), but replacing $C^{(b)}_{jk}$ with the correlation matrix
after thresholding, $D^{(b)}_{jk}$.
\begin{equation}
\label{e:ri-thresh}
r_j^{(b)} = \sum_{k=2}^N \left(D^{(b)}\right)^{-1}_{jk}D^{(b)}_{k1} \quad {\rm for\
}j={2,\cdots,N}.
\end{equation}
Having found the coupling constants $r_j^{(b)}$, one can remove the
correlations from the channel of interest using equation
(\ref{e:remove}).

There is a problem when equation (\ref{e:ri-thresh}) is applied to real
data. Because the thresholding sets entries of the correlation matrix
to zero, $D^{(b)}_{jk}$ becomes nearly singular and its inverse
$D^{-1}$ in equation (\ref{e:ri-thresh}) becomes unstable. For example,
in Figure \ref{f:correlation}, there are two channels, slow pzt and
arm2 visibility,  which are very similar to each other. When the small
correlation elements are set to zero in the correlation matrix, the two
rows corresponding to these two channels become very close to each
other, which makes $D^{(b)}_{jk}$ nearly singular.

To solve this problem, we eliminate ``redundant''channels. Consider the
eigenvalues $\lambda$ and the eigenvectors $\Lambda$ of the matrix
$D^{(b)}_{jk}$. Note that the matrix $D^{(b)}_{jk}$ is Hermitian and
positive semi-definite. Its eigenvalues are always real and
non-negative. If the matrix $D^{(b)}_{jk}$ becomes close to a singular
matrix, $D^{(b)}_{jk}$ will have an eigenvalue $\lambda_0$ which is
very close to zero. We call the corresponding eigenvector $\Lambda^0$.
Hence,
\begin{equation}
\label{e:Lambda_0}
D^{(b)}\Lambda^0=\lambda^0\Lambda^0 {\textrm{, or }}  \sum_{k=2}^N D^{(b)}_{jk}\Lambda^0_j=\lambda^0\Lambda^0_j.
\end{equation}
When $\lambda^0$ is very close to zero, the rhs of equation
(\ref{e:Lambda_0}) vanishes. This means there is at least one row in
$D$ that can be written as a linear combination of the other rows.
Because $D$ is the correlation matrix of different channels, this
implies that at least one channel is a linear combination of the other
channels. That channel is a redundant channel and gives us no useful
additional information about the environment. We can eliminate that
channel from our channel set in order to keep the correlation matrix
far from singular.  To determine the ``best'' channel to eliminate, we
consider the absolute value of elements $\left|\Lambda^0_j\right|$ in
the eigenvector $\Lambda^0$. If $\left|\Lambda^0_k\right|$ is the
maximum of all the absolute values of elements in the eigenvector
$\Lambda^0$, this means that channel $k$ makes the maximum contribution
to the null eigenvector. Hence, we remove channel $k$ from the
environmental channel set. Then, we build a new $(N-1)^2$ correlation
matrix $D$ from the remaining N-1 channels and follow the same
procedure described above until the eigenvalues are far away from
zero.

Let us summarize our method in steps:
\begin{enumerate}
\item
Threshold the correlation matrix $ C^{(b)}_{jk} $ using equation
($\ref{e:threshmatrix}$) to get a new correlation matrix  $
D^{(b)}_{jk} $.
\item
Calculate the eigenvalues $\lambda$ and the eigenvectors $\Lambda$ of
the matrix $D^{(b)}_{jk}$.
\item
Check whether there is an eigenvalue near zero. 

If not, calculate the coupling constants using equation
(\ref{e:ri-thresh}) and remove the correlations from the channel of
interest ($X$) using equation (\ref{e:remove}).

If there is an eigenvalue $\lambda^0$ which is close to zero, find the
maximum (for example $\left|\Lambda^0_k\right|$ ) of the absolute
values of elements in the corresponding eigenvector $\Lambda^0$. Then,
eliminate the corresponding channel (for example channel $k$ if
$\left|\Lambda^0_k\right|$ is the maximum) from the channel set. That
means that we eliminate the $k$'s row and $k$'th column in $
D^{(b)}_{jk}$. Then return to step 2.
\end{enumerate}

\section{General discussion of thresholding methods}
An ideal scheme of removing correlations from the channel of interest
$X$ to obtain ${\bf \bar x}$ should have the following properties:
\begin{enumerate}
\item
If any environmental channel is rescaled, i.e. ${\bf Y_j} \Rightarrow
\alpha {\bf {Y_j}}$, it does not affect the result ${\bf \bar x}$.
\item
If any environmental channel is duplicated, i.e. ${\bf Y_{N+1} = {\bf
Y_j}}$, $N \Rightarrow N+1$, it does not affect the result ${\bf \bar
x}$.
\item
If any environmental channel is duplicated by a linear combination of
other channels, i.e. ${\bf Y_{N+1}} = \alpha_2 {\bf
Y_2}+\dots+\alpha_{N} {\bf Y_{N}}$, $N \Rightarrow N+1$, then ${\bf
Y_{N+1}}$ can be removed from the set of channels without affecting the
result ${\bf \bar x}$ . Of course, property 2 is just a special case of
property 3.
\item
If the environmental channels are replaced by any linear combination of
the original channels, i.e. ${\bf {Y_i}'} = M_{ij} {\bf Y_j}$, where
$M$ is an invertible matrix, it does not affect the result ${\bf \bar
x}$.
\item
If the environmental channels are re-labeled, it does not affect the
result ${\bf \bar x}$. This is a special case of condition 4, when
$M_{ij}$ is a permutation matrix of the set $(2,\dots,N)$.
\item
If an environmental channel is Gaussian noise and independent of other
channels, then it does not  affect the final result  ${\bf \bar x}$ at
a statistically-significant level.
\end{enumerate}

If we do not do thresholding (when the number of frequency bins $F$ in
a frequency band is very large), our method has all six properties
above.  However, if we threshold using the method described in Section
\ref{s:thresh_N} (when the number of frequency bins $F$ in a frequency
band is not large enough), our method has all the properties above
except for property 4.

We also considered two other thresholding methods. The first one is to
threshold on individual channels. We check the absolute value of the
covariance coefficient between channel $j$ and channel $1$ (which is
the channel of interest $X$) in frequency band $b$, $\rho_{j1}^{(b)}$.
If $\rho_{j1}^{(b)}$ is smaller than some threshold value $\rho^*$ then
we eliminate channel $j$ from our channel set.  If $\rho_{j1}^{(b)}$ is
greater than the threshold value $\rho^*$, then we keep that channel in
our channel set. This method has all the properties above except for
property 4. Compared with the method discussed in Section
\ref{s:thresh_N}, this method is too conservative: it does not remove
all the possible contaminating noise. It is possible that one
environmental channel ${\bf {Y_j}}$ is not correlated with the channel
of interest $X$ but is correlated with another environmental channel
${\bf {Y_k}}$.  Suppose channel ${\bf {Y_k}}$ is correlated to the
channel of interest $X$, and contributes to the removal of correlated
noise from the channel of interest $X$ by equation (\ref{e:remove}). In
this situation, if we include channel ${\bf {Y_j}}$ in the channel set,
it is equivalent to the following two operations. First, we remove the
correlation between channel ${\bf {Y_j}}$ and channel ${\bf {Y_k}}$
from channel ${\bf {Y_k}}$. We call the result ${\bf {\bar Y_k}}$. Then
we remove the correlation between channel $X$ and channel ${\bf {\bar
Y_k}}$ from channel $X$. This is better than only removing the
correlation between channel ${\bf {Y_k}}$ and channel $X$ from channel
$X$ because our estimation of the correlation between ${\bf {\bar
Y_k}}$ and $X$ is better than our estimate of the correlation between
${\bf {Y_k}}$ and $X$.

Another thresholding method is to consider the eigenvectors of the
correlation matrix between the environmental channels. The correlation
matrix is diagonalized by a similarity transformation, which is a unitary
matrix $U$ made up of the eigenvectors of the correlation matrix.
\begin{equation}
L=U^{\dagger} C U
\end{equation}
Here, the matrix $L_{ij}$ is a diagonal matrix of the eigenvalues of
the correlation matrix.
\begin{equation}
L_{ij}=\cases{ \lambda_{i} & $\textrm{if}\ i=j  \textrm{, or} $  \cr
&\cr
0 & $\textrm{otherwise} \ .$} 
\end{equation}
Construct new channels ${\bf {Y_i}'}$ by ${\bf {Y_i}'} = U_{ij} {\bf
Y_j}$. These channels ${\bf {Y_i}'}$ are independent of each other
since they have vanishing correlation. Then threshold on each channel
${\bf {Y_i}'}$ individually using the method of thresholding in the
two-channel case described in Section \ref{s:uncorrelated}. However,
there is a problem with this apparently promising method. If any
environmental channel is rescaled, i.e. ${\bf Y_j} \Rightarrow \alpha
{\bf {Y_j}}$, the eigenvector of the correlation matrix is changed.
Hence the independent channels that we build ${\bf {Y_i}'} = U_{ij}
{\bf Y_j}$ are also changed. So this method does not have properties 4
and 5. One may argue that we can normalize the environmental channels
first and then diagonalize the correlation matrix by the unitary matrix
$U$. In this way $U$ is unique. However we can not find any physical
reason to use a unitary matrix to diagonalize the correlation matrix.
If we use non-unitary matrix, it is no longer unique. To demonstrate
that the non-unitary matrix is non-unique, construct a matrix $M$
\begin{equation}
M_{ij}=\cases{ 1/\sqrt{\lambda_{i}} & $\textrm{if}\ i=j  \textrm{,
  or}$
  \cr &\cr
0 & $\textrm{otherwise} \ .$} 
\end{equation}
It is obvious that $M^{\dagger}=M$, and $I=(UM)^{\dagger} C UM$ is the
diagonal identity matrix. We can arbitrarily choose another unitary
matrix $U'$.
\begin{equation}
I=U'(^{\dagger} (UM)^{\dagger} C UM) U'
 =(UMU')^{\dagger} C (UMU').
\label{e:non-uni}
\end{equation}
Let matrix $P=UMU'$. Equation (\ref{e:non-uni}) shows that $P$
diagonalizes the correlation matrix $C$ to a unit matrix $I$. Because
$M$ is non-unitary, $P$ is non-unitary. Because of the choice of $U'$
is arbitrary, $P$ is not unique. Even when only a unitary matrix is
used, there is still a problem. If an environmental channel is
duplicated, i.e. ${\bf Y_{N+1} = {\bf Y_j}}$, $N \Rightarrow N+1$, the
n eigenvectors of the correlation matrix are changed. This means this
that method does not have properties 2 and 3.

It seems difficult to find a method of thresholding which has all six
desired properties. There is a tradeoff in choosing a suitable method.
In practice, when full-scale LIGO begins operation, we expect that the
methods discussed here will provide some guidance in choosing a
suitable set of environmental signals to use in ``clean up'' and
understanding the interferometer's output. We anticipate that with some
experience and experimentation, it will not prove too difficult to
identify a set of suitable channels in different frequency bands, and
thresholds can be set based on experience and on understanding of the
instrument.

\section{Conclusion}
The methods described in this paper amount to estimating whether or not
a signal of interest is correlated with other environmental channels.
The key assumption is that the quantity being measured in the signal
channel should not have any correlations with the environment.  The
correlations are removed following a prescription that minimizes the
power in user-defined frequency bands.

We assume that the correlations with the environment are described by
linear transfer functions.  The methods used to identify and remove
these correlations are very similar to Principal Component Analysis
(PCA) carried out in frequency space.  We have used a real data set to
demonstrate that the method is both reasonable and effective.

When the full scale LIGO interferometers begin operation in the year
2000, there will be over a thousand environmental and control channels
being monitored, and the problem of identifying and eliminating the
most significant environmental contamination will be severe.  In the
end, we suspect that the methods described here will be useful in two
ways.  First, they will assist in identifying which environmental
channels are having the greatest effects on instrument performance.
The frequency dependence of these effects might be helpful in trying to
determine how they can be alleviated or eliminated.  Second, after the
most relevant set of environmental channels have been successfully
identified, these techniques should make it possible to ``clean up" the
signal, although further study will be needed to determine if this has
undesirable side effects.

\section{Acknowledgements}
We gratefully acknowledge the assistance of the LIGO project, and in
particular the assistance that we have received from Stan Whitcomb and
Fred Raab in understanding the 40-meter prototype.  This work has been
supported by NSF grants PHY-9507740 and PHY-9728704 and
Forbairt grant IC/1998/026.

\appendix
\section{probability distribution of $\rho^2$ for uncorrelated Gaussian noise}
\label{s:rho}
From equation (\ref{e:false-remove}) in Section \ref{s:uncorrelated},
we know that when the two channels are independent Gaussian random
variables, the method described in Section~\ref{s:twochannel} will
falsely remove ``correlations'' which do not exist. One method to avoid
this false dismissal of ``correlation'' is to threshold on the
coherence $\rho^2$ defined by equation (\ref{e:rho}). To set a
reasonable threshold on $\rho^2$, we need to know the probability
distribution of $\rho^2$ for the case where $\tilde X$ and $\tilde Y$
are not correlated.

To determine the probability distribution of $\rho^2$, we first
consider an $F$-dimensional complex Gaussian random variable $Z(j) \equiv
R(j) + iI(j),j\in 1\dots F$, where $R(j)$ and $I(j)$ are independent
real Gaussian random variables with vanishing mean and unit variance.
Note that in order to make the notation simpler, we introduce a new
symbol $Z$ to represent the $\tilde X$ or $\tilde Y$ in previous
Sections. The probability distributions are (subscript ``g'' means
``Gaussian'')
\begin{eqnarray}
{p_g}(R(j))&=&{1\over \sqrt {2\pi}} e^{-{R(j)^2/ 2}},\  
\textrm{and} \cr {p_g}(I(j))&=&{1\over \sqrt {2\pi}} e^{-{I(j)^2/ 2}}.
\label{e:complexarray}
\end{eqnarray} 
Define $U_Z(j) \equiv \left|Z(j)\right|^2 = R(j)^2 + I(j)^2$. The
probability distribution $p_u(U_Z(j))$ is defined by
\begin{eqnarray}        
\nonumber 
&&\int_{-\infty}^{\infty}
W(U)p_u(U)\mbox{d}U\equiv \phantom{verybigspace}\\
&&\quad \int_{-\infty}^{\infty} \int
_{-\infty}^{\infty}W\left( R^2+I^2 \right)
p_g(R)p_g(I)\mbox{d}R\mbox{d}I
\end{eqnarray} 
for any choice of function $W$. Taking $W(x)=\delta(U_Z(j)-x)$ yields
\begin{equation}
p_u(U_Z(j))= \cases{ {1\over2}e^{-{U_Z(j) / 2}} & $\textrm{for}\ 
 U_Z(j)>0\textrm{, or}$  \cr
&\cr
0 & $\textrm{  for}\ U_Z(j)<0 \ .$}
\end{equation} 
Define $U_Z \equiv \sum\limits_1^F U_Z(j)$. In the F-dimensional real
space spanned by $(U_Z(1),\dots,U_Z(N))$ the joint probability
distribution $p(U_Z(1),\dots,U_Z(F))$ is
\begin{eqnarray}
\nonumber         
&&p(U_Z(1),\dots,U_Z(F))=p_u(U_Z(1))\dots p_u(U_Z(F))\phantom{space}\\
&&\quad= 
\cases{
{\left({1\over2}\right)^n} e^{-U_Z / 2} & if $U_Z(j)>0$ for $j = 1\dots F$,\cr
&\cr
0 & $\textrm{ otherwise}$.
}
\label{e:Uz}
\end{eqnarray}
Now we calculate the probability distribution of $\rho^2$ assuming that
$\tilde X$ and $\tilde Y$ are independent $F$-dimensional complex
Gaussian random vectors. According to equation (\ref{e:rho}), the
coherence $\rho^2$ is defined by
\begin{equation}
\label{e:rho2}
\rho^2 \equiv {
\left| \left( Z_1, Z_2 \right) \right|^2
\over
\left( Z_1, Z_1 \right)  \left( Z_2, Z_2 \right) 
},
\end{equation}
where $Z_1$ and $Z_2$ are $F$-dimensional complex vectors. Without loss
of generality, we assume $Z_1$ and $Z_2$ both have unit norm, or
$\left(Z_1,Z_1\right)=\left(Z_2,Z_2\right)=1$, so $U_{Z_1}=U_{Z_2}=1$ .
Because equation (\ref{e:rho2}) is rotationally invariant, we can also
assume   $Z_1(1)=1$ and $Z_1(j)=0$ for $j=2,\dots F$. Then,
\begin{equation}
\rho^2 = \left|Z_2(1)\right|^2=U_{Z_2}(1).
\end{equation}
Thus, the probability distribution of $\rho^2$ is equal to the
probability distribution of $U_Z(1)$ given that $U_Z=1$, where $Z(j)$ is
an $F$-dimensional complex random variable with probability distribution
given by equation (\ref{e:complexarray}).
\begin{equation}
p(\rho^2) = p(U_Z(1)|U_Z=1)|_{U_Z(1)=\rho^2},
\end{equation}
and the cumulative probability distribution
\begin{equation}
p({\rho^2}>{\rho_0^2})=p({U_Z(1)}>\mu |U_Z=1)|_{\mu=\rho_0^2}.
\end{equation}

It will be easier to first determine the cumulative probability
distribution $p({\rho^2}>{\rho_0^2})$, or
\begin{equation}
p({U_Z(1)}>\mu |U_Z=1)=\int^1_{\mu} p(U_Z(1)|U_Z=1)\mbox{d}U_Z(1).
\end{equation}
Note that to normalize this probability distribution requires
${p({U_Z(1)}>\mu |U_Z=1)|_{\mu=0}}=1$. In the $F$-dimensional real space
spanned by $(U_Z(1),\dots,U_Z(F))$, the condition $U_Z=1$ defines part
of an ($F-1$)-dimensional plane $P_0$ defined by 
\begin{eqnarray}
P_0:
\cases{
&$U_z(1)+\dots+U_z(F)=1 $\cr
&\cr
&$U_z(j)>0 {\textrm{ \ for }}j \in 1 \dots F$.} 
\label{e:allarea}
\end{eqnarray} 
The region ($U_Z(1)>\mu$ and $ U_Z=1$) defines a part of an
($F-1$)-dimensional plane $P_\mu$ given by
\begin{eqnarray}
P_{\mu}: 
\cases{
&$U_z(1)+\dots+U_z(F)=1 $\cr &\cr
&$U_z(1)>\mu $\cr &\cr
&$U_z(j)>0{\textrm{ \ for }}j \in 2 \dots F $. 
}
\end{eqnarray} 
Note that $P_\mu$ is a subset of $P_0$. From equation (\ref{e:Uz}), we
can see that  $p(U_Z(1),\dots,U_Z(F))$ is a constant for any given
$U_Z$. In our case, $U_Z=1$. Hence $p({U_Z(1)}>\mu|U_Z=1)$ is just the
ratio between the $(F-1)$-volume of $P_\mu$ and the $(F-1)$-volume of
$P_0$. To help calculate the volume of $P_\mu$, we can translate the
coordinates $(U_Z(1),\dots,U_Z(F))$  so that the origin moves to the
point $(\mu,0,\dots,0)$:
\begin{eqnarray}
{\textrm{ new coordinates:}}
\cases{
&$U_z(1)'= U_z(1)-\mu $\cr
&\cr
&$U_z(j)'= U_z(j){\textrm{ \ for }}j \in 2 \dots F.$ }
\end{eqnarray} 
In the $F$-dimensional real space spanned by the coordinates 
$(U_Z(1)',\dots,U_Z(F)')$, 
the plane  $P_{\mu}$ is given by
\begin{eqnarray}
P_{\mu}: \cases{
&$U_z(1)'+\dots+U_z(F)'=1-\mu $\cr &\cr
&$U_z(j)'>0{\textrm{ \ for }}j \in 1 \dots F$. 
}
\label{e:toparea}
\end{eqnarray} 
Comparing equation (\ref{e:allarea}) and equation (\ref{e:toparea}), we
see that $P_\mu$ and $P_0$ are rescaled versions of each other, and the
linear dimension of $P_\mu$ equals $(1-\mu)$ times the linear dimension
of $P_0$. Hence,
\begin{eqnarray}
\nonumber
p({U_Z(1)}>\mu|U_Z=1)&=&{(F-1){\textrm{-volume of}}\  {P_\mu} \over
(F-1){\textrm{-volume of}}\ {P_0}}\\
 &=& {\left(1-\mu\right)}^{F-1}
\end{eqnarray}
Thus the cumulative distribution
\begin{equation}
p({\rho^2}>{\rho_0^2})={\left(1-{\rho_0^2}\right)}^{F-1}
\label{e:cumulative}
\end{equation}
Note that $p({\rho^2}>{\rho_0^2})|_{\rho_0^2=0}=1$. Hence this
probability distribution is correctly normalized.
 
Taking the derivative of equation (\ref{e:cumulative}) to get the
differential probability distribution function yields
\begin{eqnarray}
p(\rho^2)&=&-{{\mbox{d}p({\rho^2}>{\rho_0^2})}
\over{\mbox{d}{\rho_0^2}}}
|_{{\rho_0^2}={\rho^2}}\cr
&=&(F-1){\left(1-\rho^2\right)}^{F-2}.
\end{eqnarray}
Hence,
\begin{equation}
p(\rho^2)=(F-1){\left(1-\rho^2\right)}^{F-2}.
\end{equation}
The expected value of ${\rho^2}$ is
\begin{equation}
<\rho^2>=\int \rho^2 p(\rho^2)\mbox{d} \rho^2 = {1 \over F}.
\end{equation}
Now we can return to the problem identified at the beginning
of this section.

The aim was to avoid false dismissal of non-existing ``correlation'' by
setting a reasonable threshold on the coherence $\rho^2$ (defined by
equation (\ref{e:rho})) between channels $\tilde X$ and $\tilde Y$. If
$\rho^2$ is greater than the threshold ${\rho^*}^2$, we conclude that
the correlation between $\tilde X$ and $\tilde Y$ is present at a
statistically-significant level, and remove the correlation using
method described in Section~\ref{s:twochannel}. If not, we leave
channel $\tilde X$ unchanged.  According to equation
(\ref{e:cumulative}), when two channels are just independent Gaussian
random variables, the probability of incorrectly removing
``correlation'' between them is given by
$\left(1-(\rho^*)^2\right)^{F-1}$. For example, when $F=128$ and
$(\rho^*)^2=10/F$, the probability of incorrectly removing
``correlation'' is $\approx 3 \times 10^{-5}$.


\bibliographystyle{IEEE}

\end{document}